\documentstyle[12pt,a4]{article}

\begin{document}
 \newcommand{\PRL}{{\em Phys. Rev. Lett. }}
 \newcommand{\PRA}{{\em Phys. Rev. }}
 \newcommand{\SOV}{{\em Sov. Phys. JETP }}
 \newcommand{\NP}{{\em Nucl. Phys. }}
 \newcommand{\JPB}{{\em Jour. of Phys. }}
 \newcommand{\IJMP}{{\em Int. Jour. of Mod. Phys.  }}
\newcommand{\beq}{\begin{equation}}
\newcommand{\eeq}{\end{equation}}
\newcommand{\bea}{\begin{eqnarray}}
\newcommand{\eea}{\end{eqnarray}}
\newcommand{\hs}{\hspace{0.1cm}}
\newcommand{\spz}{\hspace{0.7cm}}
 \newcommand{\reset}{  \setcounter{equation}{0}  }
\newcommand{\th}{\theta}
\newcommand{\thb}{\bar{\theta}}
\newcommand{\ib}{ \bar{ \imath}}
\newcommand{\jb}{ \bar{ \jmath}}
\newcommand{\etab}{ \bar{ \eta}}
\newcommand{\kb}{ \bar{ k}}
\newcommand{\s}{\sigma}
\newcommand{\be}{\beta}
\newcommand{\Mrr}{I \!\! R} 
\newcommand{\xh}{ \hat{ x}}
\newcommand{\zb}{\bar{z}}
\newcommand{\zt}{\tilde{\zeta}}
\newcommand{\et}{\tilde{\zeta}}
\newcommand{\k}{\kappa}
\newcommand{\PL}{{\em Phys. Lett. }}
 \newcommand{\JMP}{{\em Jour. of Math. Phys.  }}
\newcommand{\CMP}{{\em Comm. Math. Phys.  }}
\setcounter{page}{0}
\topmargin 0pt
\oddsidemargin 5mm
\renewcommand{\thefootnote}{\fnsymbol{footnote}}
\newpage
\setcounter{page}{0}
\begin{titlepage}
\begin{flushright}
Berlin Sfb288 Preprint  \\
quant-ph/9609011\\
2-nd revised version \\
\end{flushright}
\vspace{0.5cm}
\begin{center}
{\Large {\bf On the Influence of Pulse Shapes on Ionization Probability} }

\vspace{1.8cm}
{\large C. Figueira de Morisson Faria,$^{\dag}$  A. Fring$^{*}$ 

and R. Schrader$^{*}$     }\footnote{e-mail addresses: Faria@mbi.fta-berlin.de,

Fring@physik.fu-berlin.de,

Schrader@physik.fu-berlin.de}

\vspace{0.5cm}
{\em $*$ Institut f\"ur Theoretische Physik\\
Freie Universit\"at Berlin, Arnimallee 14, D-14195 Berlin, Germany\\
${\dag}$ Max-Born-Institut, Rudower Chaussee 6, D-12474 Berlin, Germany}

\end{center}
\vspace{1.2cm}
 
\renewcommand{\thefootnote}{\arabic{footnote}}
\setcounter{footnote}{0}

\begin{abstract}
We investigate analytical expressions for the upper and lower bounds
for the ionization probability through ultra-intense shortly pulsed 
laser radiation. We take several different pulse shapes into account,
including in particular those with a smooth adiabatic turn-on and 
turn-off. For all situations for which our bounds are applicable 
we do not find any evidence for bound-state stabilization.
\end{abstract}
\vspace{.3cm}
\centerline{August 1996}
\end{titlepage}
\newpage

\renewcommand{\theequation}{\mbox{\arabic{section}.\arabic{equation}}}
\setcounter{equation}{0}

\section{Introduction}
The computation of ionization rates or probabilities of atoms through 
low intensity ($ I << 3.5 \times 10^{16} W/cm^2$ ) laser radiation  
can be carried out
successfully using perturbation theory around the solution of the
Schr\"odinger equation without the presence of the laser fields \cite{Lam}. 
With the advance of laser technology, nowadays intensities of up to 
($10^{19} W/cm^2$) are 
possible  and pulses may be reduced to a
duration of ($\tau \sim  10^{-15}s$),
\footnote{For a review and the experimental realization of such pulses
see for instance \cite{Wil}.}   the region of
validity of the above method is left. The new regime is usually tackled
by perturbative methods around the Gordon-Volkov solution \cite{GordonVolk}
of the Schr\"odinger equation 
\cite{Keld,GeltT,Reiss,Gelt1,Grobe,Gelt2,Gelt3}, 
fully numerical solutions 
of the Schr\"odinger equation \cite{Col,Bard,LaG,SuEb,Burnett,Dorr}, 
Floquet solution
\cite{Floquet,Pot,Faisal}, 
high frequency approximations \cite{HFA} or analogies
to classical dynamical systems \cite{classical}. All these methods have its
drawbacks.
The most surprising outcome of the analysis of the high-intensity
region for short pulses (the pulse length is smaller than  1 ps)
is the finding by the majority of the atomic-physics community 
(see  \cite{BRK,KulS,KrainDel,Gelt4} and references therein) 
of so-called atomic stabilization. This means that the 
probability of ionization by a pulse of laser radiation, which for
low intensities increases with increasing intensities reaches some 
sort of maximum at high intensities and commences to decrease until
ionization is almost totally suppressed. This picture is very 
counterintuitive and doubts on the existence of this phenomenon have been 
raised by several authors \cite{Gelt2,Chen,Krain,Gelt1,Gelt3}, 
who do not find evidence for it in
their computations. So far no support is given to either side by
experimentalists.\footnote{Experimental evidence for some sort of 
stabilization is
given in \cite{expe}, but these experiments deal with intensities
of $10^{13}W/cm^2$, which is not the "ultra-intense" regime for which
the theoretical predictions are made.}
For reviews on the subject we refer to \cite{BRK,KulS,KrainDel,Gelt4}.
 
Since all of the above methods involve a high degree of numerical
analysis, which are difficult to be verified by third parties, it is
extremely desirable to reach some form of analytical understanding.
In \cite{EKS,KS1,KS2,FKS} we derived analytical expressions for  upper and 
lower bounds for  the ionization probability, meaning that the
ionization probability is certainly lower or higher, respectively,
than these values. The lower bound in particular may be employed to 
investigate the possibility of  stabilization for an atomic bound state.
In \cite{FKS} we analyzed the hydrogen atom and found that for increasing
intensities the lower bound also increases and hence that the existence
of atomic stabilization can be excluded in the sense that the ionization
probability tends to one. The shortcoming of our 
previous analysis  \cite{FKS} is, 
that definite conclusions concerning the above
question may only be reached for extremely short pulses ($ \tau < 1$ a.u.),
which are experimentally unrealistic.
In the present article we analyze these bounds in further detail
and demonstrate that atomic stabilization can also be excluded for longer 
pulses.

Some authors \cite{Grobe,SuEb} put forward the claim that in order 
to ``observe"
atomic stabilization one requires pulses which are switched on,
sometimes also off, smoothly. This seems very surprising since
stabilization is supposed to be a phenomenon specific to high
intensities and with these type of pulses  emphasis is just put on
the importance
of the low intensity regime. It further appears that among the 
authors who put forward these claims, it is not commonly agreed
upon, whether one should associate these pulse shapes to the laser
field or to the associated vector potential. 
We did not find a proper and convincing
physical explanation why such pulses should produce so surprising effects
in the literature.
Geltman \cite{Gelt3}  and also Chen and Bernstein \cite{Chen} do not find
evidence for stabilization for these type of pulses with smooth 
and turn on (and  off) of the laser field.

In order to address also the validity of these claims in our framework 
we extend in
the present paper our previous analysis to various type of pulses 
commonly employed in the literature in this context
and 
investigate also the effects different frequencies might have. Once
more we conclude that our arguments do not support atomic stabilization.    

Our manuscript is organized as follows: In section 2. we briefly recall
the principle of our argumentation and our previous expressions 
for the upper and lower bounds for the ionization
probability and discuss them in more detail for the hydrogen atom. 
We then turn to an  
analysis for specific pulses. In section 3. we state our conclusions. 
In the appendix we present the explicit computation
for the Hilbert space norm of the difference
 of the potential in the 
Kramers-Henneberger frame and the one in the laboratory frame for 
any bound state. 
\section{The upper and lower bounds}
For the convenience of the reader we commence by summarizing briefly the 
main principle of our argument. Instead of calculating exact ionization 
probabilities we compute upper and lower bounds for them, meaning that the
exact values are always greater or smaller, respectively. We then vary these
bounds with respect to the intensity of the laser field and study their 
behaviour. If the lower bound tends to one with increasing intensity, we can
infer that stabilization is definitely excluded. On the other hand, if 
the upper bound tends to zero for increasing intensities, we would conclude
that stabilization is present. In case the lower bound increases, but
remains below one, we only take this as an indication for a general type
of behaviour and interpret it as not providing any evidence for stabilization,
but we can not definitely exclude its existence. In case the lower bound 
becomes negative or the upper bound greater than one, our expressions
obviously do not allow any conclusion.  
\par
The non-relativistic quantum mechanical description of a 
system with potential $V$ in the presence of  linearly
polarized laser radiation is given by the Schr\"odinger equation involving 
the Stark Hamiltonian
\beq
i\frac{\partial\psi(\vec{x},t)}{\partial t}
\;= \;\left(-\frac{\Delta}{2}+V+ z\cdot E(t)\right)\psi(\vec{x},t)  
\; = \; H(t) \psi(\vec{x},t)  .
\eeq
For high, but not relativistic, intensities the laser field may be approximated
classically. We furthermore assume the dipole approximation. In the
following we will always use atomic units $ \hbar=e=m_e=c \cdot \alpha=1$.
For a general time dependent Hamiltonian $H(t)$   
the ionization probability of a  normalized bound state $\psi$ is defined 
\cite{Reiss,EKS} as
\beq
P(\psi) \;= \; \|({\bf 1}-{\cal P}_{+ })S\psi\|^2 \;=\;
  1 -  \| {\cal P}_{+ }S\psi\|^2 \;\;\; .
\eeq
The gauge invariance of this expression was discussed in \cite{FKS}.
Here $ \| \psi \| $ denotes as usual the Hilbert space norm, i.e. 
$\| \psi \|^2 = \langle \psi, \psi  \rangle = \int 
|\psi(\vec{x})|^2 d^3x $. We always assume that 
$H_{\pm} = \lim\limits_{t \rightarrow \pm \infty} H(t)$
exists and $\psi$ is then understood to be a bound state of $H_-$. 
${\cal P}_{+}$ and ${\cal P}_{-}$  denote the projectors onto 
the space spanned by the bound states of $H_+$ and $H_-$, respectively 
and  S is the unitary ``scattering matrix''
\beq
S=\lim_{t_{\pm} \rightarrow  \pm \infty} \exp( it_+ H_+ ) \cdot 
U(t_+,t_-) \cdot  \exp (-it_- H_- ) \;\; .
\eeq
Here the unitary time evolution operator $U(t_+,t_-)$ for $H(t)$, 
brings a state from time $t_-$ to $t_+$. 
Note that by definition  $ 0 \leq P(\psi) \leq 1$.
Employing methods of functional 
analysis
we derived in \cite{EKS,KS1,KS2,FKS} several analytical expressions 
by which the possible values for the ionization probability may be
restricted. We emphasize once more that these  expressions are not to 
be confused with exact computations of ionization probabilities. 
We recall here the formula for the upper 
\begin{equation}\label{upo}
P_u(\psi)^{1\over 2} = \int\limits^\tau_0 \| (V(\vec{x} -c(t)e_z)
-V(\vec{x}))\psi\|dt
+ |c(\tau)|~\| p_z\psi\| +|b(\tau)|~\| z\psi\|
\end{equation}
and  the lower bound 
\begin{eqnarray}\label{lowo}
P_l(\psi) =  1&-&\Bigg\{ \int\limits^\tau_0\| 
(V(\vec{x} -c(t)e_z)-V(\vec{x}))\psi\|dt\\&&+ \frac{2}{2E+ b(\tau)^2} \| 
(V(\vec{x} -c(\tau)e_z)-V(\vec{x}))\psi\|+\frac{2 |b(\tau)|}{2 E+ b(\tau)^2} 
\|p_z\psi\|\Bigg\}^2 \;\; , \nonumber
\end{eqnarray}
which were deduced in \cite{FKS}.
$e_z$ is the unit vector in the z-direction.  
Here we use the notation
\beq 
b(t) := \int^t_0 E(s) ds \qquad \qquad
c(t) := \int^t_0 b(s) ds \;\; ,
\eeq
for the total classical momentum transfer and the total classical
displacement, respectively.
Note that for the vector potential in the z-direction 
we have $A(t) = - \frac{1}{c} b(t)+ const$.
It is important to recall that the expression for the lower bound 
is only valid if the classical energy transfer is larger than the 
ionization energy of the bound state, i.e.  $\frac{1}{2} b^2(\tau) > - E$.
Our bounds hold for all Kato small potentials.\footnote{Potentials 
are called Kato small if for arbitrary there      
$0< a < 1$ there is a constant  $b < \infty$, such that
$\|V\psi\|\le a\|-\Delta\psi\|+b\|\psi\|$ holds 
for all $\psi$ in the domain ${\cal D}(H_0)$ of $H_0= -\Delta/2 $,
see for instance \cite{CFKS,RS}. }
In particular the Coulomb potential and its modifications, which are very 
often  employed  in numerical computations,
such as smoothed or screened Coulomb potentials, are Kato small. However,
the delta-potential, which is widely used in toy-model computations 
because of its nice property to possess only one bound state, is not a 
Kato potential.

In the following we will consider a realistic example and  take the 
potential $V$ to be the Coulomb potential
and concentrate our discussion on the hydrogen atom. In this case it is well
known that the binding energy  for a state $\psi_{nlm}$ is
$E_n = -\frac{1}{2n^2}$, $ \| p_z \psi_{n00} \|^2 =
\frac{1}{3 n^2}$ and
$ \| z \psi_{n00} \|^2 = \frac{1}{3} \langle \psi_{n00} | r^2 | 
\psi_{n00}  \rangle = \frac{n^2}{6} (5 n^2 +1) $ 
(see for instance \cite{BSLL}). 
We will employ these relations below. 
In \cite{FKS} it was shown, that the Hilbert space 
norm of the difference of the potential in the Kramers-Henneberger frame
\cite{K,H}
and in the laboratory frame applied to the state $\psi$
\beq
N(\vec{y},\psi) \; := \; \| (V(\vec{x} - \vec{y})-V(\vec{x}))\psi\|
\eeq 
is bounded by 2 when  $\psi = \psi_{100}$ for arbitrary  $\vec{y} = c e_z$.
We shall now investigate in more detail how this function depends on
c. In order to simplify notations we ignore in the following the 
explicit mentioning of $e_z$.
In the appendix we present a detailed computation, where  we obtain
\beq \label{HNorm}
N^2(c,\psi_{100}) \; = \; 2 + (1+|c|^{-1}) e^{-|c|} Ei\left( |c| \right)
+ (1-|c|^{-1}) e^{|c|} Ei\left( -|c| \right)
+\frac{2}{|c|}\left( e^{-2 |c|} -1 \right).
\eeq
Here $Ei(x)$ denotes the exponential integral function, given by the  
principal value of the integral
\beq
Ei(x) \; = \; - \int\limits_{-x}^{\infty} \frac{e^{-t}}{t} \; dt 
\qquad \hbox{for}\quad x>0 \;\; .
\eeq
Considering now the asymptotic of $ N$, we obtain as expected $
\lim\limits_{ c \rightarrow 0} N =0 $ and $ \lim\limits_{ c
\rightarrow \infty} N =\sqrt{2} $. Noting further that $ N$
is a monotonically increasing function of $c$, (one may easily compute its
derivatives w.r.t. c, but we refer here only to the plot of this function
in figure 1), 
it follows that our previous \cite{FKS}
estimate may in fact be improved to $N(c,\psi_{100}) \leq \sqrt{2}$. The
important thing to notice is, that since $N(c,\psi_{100})$ is an overall
increasing function of $c$, it  therefore also increases as a
function of the field strength. The last term in the bracket of the
lower  bound $P_l(\psi)$ is a
decreasing function of the field strength, while the second term does not
have an obvious behaviour. Hence if the first term dominates the whole
expression in the bracket, thus leading to a decrease of $P_l(\psi)$, 
one has in principle the possibility of stabilization. 
We now investigate several pulse shapes for the possibility of 
such a behaviour and analyze the expressions
\bea \label{low}
P_l(\psi_{100}) &=& 1 - \Bigg\{ \int\limits^\tau_0 N(c(t),\psi_{100}) dt  
+ \frac{2 N(c(\tau),\psi_{100})}{b(\tau)^2 -1  }  
+\frac{2}{\sqrt{3}} \frac{|b(\tau)|}{b(\tau)^2 - 1}  \Bigg\}^2 \;\;\;\;\;\; \\
P_u(\psi_{100}) &=& \left\{  \int\limits^\tau_0 N(c(t),\psi_{100}) dt 
+ \frac{|c(\tau)|}{\sqrt{3}} + | b(\tau)|   \right\}^2 \label{up}
\;\;\; .
\eea
Here we have simply inserted the explicit values for $E_1$,
$ \| z \psi_{100} \|$ and $ \| p_z \psi_{100} \|$ into
(\ref{upo}) and (\ref{lowo}), and understand $N(c,\psi_{100})$
to be given by the analytical expression (\ref{HNorm}). 
The formulae presented in the
appendix allow in principle also the computation of  
$N(c,\psi_{nlm})$ for different values of $n,l$ and $m$. However, for 
$l \neq 0$ the sum over the Clebsch-Gordan coefficients becomes
more complicated and 
due to the presence of the  Laguerre polynomial of degree n
in the radial wave-function $R_{nl}$  this becomes a rather complex 
analytical computation. We will therefore
be content with a weaker analytical  estimate here. 
In fact, we have 
\beq
N^2(c(t),\psi_{n00})  \; \leq \; 2 \langle \psi_{n00},V(\vec{x})^2 
\psi_{n00} \rangle \; =\; \frac{4}{n^3} \;\; .
\eeq
In the appendix of \cite{FKS} this statement was proven for $n=1$. The
general proof for arbitrary n may be carried out exactly along the same 
line. Therefore, 
we obtain the following new  upper and lower bounds 
\bea
\label{lowcrude} 
P_{lw}(\psi_{n00}) &=& 1 - \Bigg\{
\frac{2}{n^{3/2} }  \tau + \frac{4}{b(\tau)^2 -1/n^2 } 
\frac{1}{n^{3/2}}   
+\frac{1}{n\sqrt{3}}\frac{2|b(\tau)|}{b(\tau)^2-1/n^2}\Bigg\}^2 
\label{lowwcrude} \\
P_{uw}(\psi_{n00}) &=&  \Bigg\{
\frac{2}{n^{3/2} }  \tau + \frac{ |c(\tau)| }{n \sqrt{3}} 
+ n \sqrt{ \frac{5 n^2 +1}{6} }   |b(\tau)| \Bigg\}^2, \label{upcrude}
\eea
which are weaker than (\ref{up}) and (\ref{low}), in the the sense that
\beq
P_{lw}(\psi_{n00}) \leq P_{l}(\psi_{n00}) \leq
P(\psi_{n00}) \leq P_{u}(\psi_{n00}) \leq P_{uw}(\psi_{n00}) \;\;\; .
\label{ndep}
\eeq
In order for (\ref{lowcrude}) to be valid we now have to have 
$ b(\tau)^2 > \frac{1}{n^2}$.
We will now turn to a detailed analysis of these bounds by looking at
different pulses. Our main purpose in the present manuscript for considering
states of the type $\psi_{nlm}$ with $n \neq 0$ is to extend our discussion
to pulses with longer duration, see also section 2.3. The reason that longer
pulse durations are accesible for states with higher n is the n-dependence in 
estimate (\ref{ndep}) and its effect in (\ref{upcrude}) and
(\ref{lowwcrude}).
\subsection{Static Field}
This is the  simplest case, but still instructive to
investigate since it already contains the general feature which we
will observe for more complicated pulses. It is furthermore important
to study, because it may be viewed  as the background which is present in most
experimental setups, before more complicated pulses can be generated.
For a static field of intensity $I = E_0^2$ we trivially have 
\beq 
E(t) =
E_0 \qquad b(t) =E_0 t \qquad c(t) = \frac{E_0 t^2}{2} 
\eeq 
for $ 0 \leq t \leq \tau$. Inserting these functions 
into (\ref{low}) we may
easily compute the upper and lower bound. Here the one dimensional 
integrals over
time, appearing in (\ref{up}) and (\ref{low}) were carried 
out numerically. The result is presented in figure 2, which shows
that a bound for higher intensities always corresponds to a 
higher ionization probability. The overall qualitative behaviour
clearly indicates that for increasing field strength the
ionization probability also increases and tends to one.
In particular lines for different intensities never cross each other. 
Surely the shown pulse lengths are too short to be realistic and we 
will indicate below how to obtain situations in which conclusive
statements may be drawn concerning longer pulse durations. 
In the following we will always encounter the same qualitative 
behaviour.  
\subsection{Linearly polarized monochromatic light (LPML)}
Now we have 
\beq E(t) = E_0 \sin(\omega t) \qquad b(t) = \frac{2
E_0}{\omega} \sin^2 \left(\frac{\omega t}{2} \right) \qquad c(t) =
\frac{E_0}{\omega^2} \left( \omega t - \sin( \omega t ) \right) 
\eeq 
for $ 0 \leq t \leq \tau$. 
The result of the computation which employs
these functions in order to compute (\ref{low}) and (\ref{up}) 
is illustrated in
figure 3. Once again our bounds indicate  that for increasing field
strength the ionization probability also increases. Keeping the field
strength fixed at $E_0 = 2$ a.u., a comparison between the case for 
$\omega = 0.4$ and
$\omega = 4$ shows (figure 4), as expected, the  
lower  bounds for the ionization 
probability to be  decreasing functions of the frequency. The peak on the 
left, which seems to contradict this statement for that region, is only
due to the fact that the expression for the lower bound is not valid for
$\omega = 0.4$ in that regime. Clearly, this is
not meant by stabilization, since for this to happen we require  fixed 
frequencies
and we have to analyze the behaviour for  varying field strength. 
The 
claim \cite{SuEb,HFA} is that in general very high frequencies 
are required for this
phenomenon to emerge. Our analysis does not support stabilization 
for any frequency. As mentioned above, the shortcoming of the analysis 
of the bounds $P_u(\psi_{100})$ and $P_l(\psi_{100})$ is that   
we only see an effect for 
times smaller than one atomic unit. figure 4 and figure 5 also show that 
by considering $P(\psi_{n00})$ for 
higher values of $n$ our expressions allow also conclusions for
longer pulse durations.  For the reasons
mentioned above, in this analysis we employed the slightly
weaker  bounds
(\ref{upcrude}) and  (\ref{lowcrude}). 
\subsection{LPML with a trapezoidal enveloping function}
We now turn to the simplest case of a pulse which is adiabatically
switched on and off. These type of pulses are of special interest since many
authors claim \cite{SuEb,Grobe} 
that stabilization only occurs in these cases.  
We consider a pulse of duration $\tau_0$ which has linear turn-on and 
turn-off ramps of length $T$. Then
\bea 
E(t) &=&  E_0 \sin( \omega t) \left\{ \begin{array}{ll}
\frac{t}{T} & \hbox{for}\qquad 0 \leq t \leq T \\ 
1  & \hbox{for} \qquad  T < t < (\tau_0 -T)  \\ 
\frac{(\tau_0 - t)}{T} & \hbox{for} \qquad  (\tau_0 -T) \leq  t \leq \tau_0 
\end{array} \right.  \\
b(\tau_0) &=& \frac{E_0}{ \omega^2 T} 
\left\{\sin(\omega T) - \sin(\omega \tau_0) + \sin(\omega
(\tau_0 - T)) \right\}  \\
c(\tau_0) &=& \frac{E_0}{  \omega^3  T} 
\Biggl(2 - 2 \cos(\omega T) + 2 \cos(\omega \tau_0) - 2 \cos(\omega (\tau_0-T)) 
  \nonumber \\ & &\;\;\;\;\;\;\;
      -   \omega T \sin(\omega T) + \omega \tau_0 \sin(\omega T) + 
          \omega T \sin(\omega (\tau_0 -T)) \Biggr) \;\; .
\eea
The expressions for $b(t)$ and $c(t)$ are rather messy and will not be
reported here since we only analyze the weaker bounds. Notice that
now, in contrast to the previous cases, both $b(\tau_0)$ and  $c(\tau_0)$  
may become zero for certain pulse durations and ramps.
We shall comment on this situation in section 3. 
We choose the
ramps to be of the form $T = \left( m + \frac{1}{4}\right) 
\frac{2 \pi}{\omega}$  (m being an integer) for the lower  and 
$T = \left( m + \frac{1}{2}\right) \frac{2 \pi}{\omega}$
for the upper bound. Our lower bound does not permit the analysis
of half cycles since then $b(\tau_0) =0$. 
The results are shown in
figure 6 and 7, which both do not show any evidence for stabilization.
They further indicate 
that a decrease in the slopes of the ramps with fixed pulse duration,
leads to a smaller ionization probability. Once more (we do not present
a figure for this, since one may also see this from the analytical
expressions), 
an increase in the 
frequency leads to a  decrease in the lower bound of  
the ionization probability for fixed field strength. 

\subsection{LPML with a sine-squared enveloping function}
Here we consider
\bea E(t) &=& E_0 \; \sin^2\left( \Omega t\right) \sin(\omega t) \\
b(t) &=& { \frac{E_0}{ 16\,{\it \omega}\,{{{\it \Omega}}^2} -4\,{{{\it
\omega}}^3}} } \,\Biggr( 8\,{{{\it \Omega}}^2} + 2\,{{{\it
\omega}}^2}\,\cos ({\it \omega}\,t) - 8\,{{{\it \Omega}}^2}\,\cos
({\it \omega}\,t) \nonumber \\ & & - {{{\it \omega}}^2}\,\cos (\left(
{\it \omega} - 2\,{\it \Omega} \right) \,t) - 2\,{\it \omega}\,{\it
\Omega}\, \cos (\left( {\it \omega} - 2\,{\it \Omega} \right) \,t)
\nonumber \\ & & - {{{\it \omega}}^2}\,\cos (\left( {\it \omega} +
2\,{\it \Omega} \right) \,t) + 2\,{\it \omega}\,{\it \Omega}\, \cos
(\left( {\it \omega} + 2\,{\it \Omega} \right) \,t) \Biggl) \\ 
c(t) &=& {\frac{E_0}{4\,{{{\it \omega}}^2}\,{{\left( {\it \omega} - 2\,{\it
\Omega} \right) }^2}\, {{\left( {\it \omega} + 2\,{\it \Omega} \right)
}^2}}} \,\Biggl( -8\,{{{\it \omega}}^3}\,{{{\it \Omega}}^2}\,t +
32\,{\it \omega}\,{{{\it \Omega}}^4}\,t - 2\,{{{\it \omega}}^4}\,\sin
({\it \omega}\,t) \nonumber \\ 
& &+ 16\,{{{\it \omega}}^2}\,{{{\it
\Omega}}^2}\, \sin ({\it \omega}\,t) - 32\,{{{\it \Omega}}^4}\,\sin
({\it \omega}\,t) - {{{\it \omega}}^4}\,\sin (\left( 
2\,{\it \Omega}  -{\it \omega} \right) \,t) \nonumber \\ 
& & - 4\,{{{\it
\omega}}^3}\,{\it \Omega}\, \sin (\left(  2\,{\it
\Omega}  -{\it \omega} \right) \,t) 
- 4\,{{{\it \omega}}^2}\,{{{\it \Omega}}^2}\,
\sin (\left( 2\,{\it \Omega} - {\it \omega}   \right) \,t) \nonumber \\
& & + {{{\it \omega}}^4}\,\sin (\left( {\it \omega} + 2\,{\it \Omega}
\right) \,t) - 4\,{{{\it \omega}}^3}\,{\it \Omega}\, \sin (\left( {\it
\omega} + 2\,{\it \Omega} \right) \,t) \nonumber \\ & & + 4\,{{{\it
\omega}}^2}\,{{{\it \Omega}}^2}\, \sin (\left( {\it \omega} + 2\,{\it
\Omega} \right) \,t) \Biggr) \eea
for $ 0 \leq t \leq \tau$. At first sight it appears that both $b(t)$ 
and $c(t)$ are  singular at $\omega = \pm 2 \Omega$, which of course is 
not the case since both functions are bounded as one may easily derive. With
the help of the Schwarz inequality it follows that always 
$|b(t)| \leq t^{ \frac{1}{2} } \|E \| $
and $|c(t)| \leq \frac{1}{2}  t^{ \frac{3}{2} } \|E \| $.
We first investigate the
situation in which this pulse is switched on smoothly but turned off
abruptly. Figure 8 shows that the bounds become nontrivial for times
larger than one atomic unit in the same fashion as in the previous
cases by considering $P_l(\psi_{n00})$ 
for higher values of n. Figure 9 shows that also
in this case the ionization probability tends to one and no sign
for stabilization is found. Figure 10 shows the lower bound in which
the pulse length is taken to be a half cycle of the enveloping function.
Once more it indicates increasing ionization probability with 
increasing field strength and also for increasing values for $n$. 
Following now Geltman  \cite{Gelt3} and Su et al. \cite{SuEb} 
we employ the sine-square only for
the turn-on and off and include a plateau region into the pulse shape.
Then
\bea 
E(t) &=&  E_0 \sin( \omega t) \left\{ \begin{array}{ll}
\sin^2\left(\frac{\pi t}{2 T} \right)   & \hbox{for}\qquad 0 \leq t \leq T \\ 
1  & \hbox{for} \qquad  T < t < (\tau_0 -T)  \\ 
\sin^2\left(\frac{\pi (\tau_0 -t)}{2 T} \right)& \hbox{for} \qquad  
(\tau_0 -T) \leq  t \leq \tau_0 
\end{array} \right.  \\
b(\tau_0) &=& \frac
{E_0\,{{\pi }^2} \,\left( 1 + \cos ({\it \omega}\,T) - 
       \cos ({\it \omega}\,\left( T - {\it \tau_0} \right) ) - 
       \cos ({\it \omega}\,{\it \tau_0}) \right) } 
   {2\,{\it \omega}\,{{\pi }^2} - 2\,{{{\it \omega}}^3}\,{T^2}} \\
c(\tau_0) &=& \frac 
{E_0\,{{\pi }^2}{2\,{{{\it \omega}}^2}\,}}
{{\left( \pi^2  - {\it \omega^2}\,T^2 \right) }^2}
       \,\Bigl( {\it \omega}\,{{\pi }^2}\,{\it \tau_0} - 
       {{{\it \omega}}^3}\,{T^2}\,{\it \tau_0} - 
       {\it \omega}\,{{\pi }^2}\,T\,\cos ({\it \omega}\,T) + 
       {{{\it \omega}}^3}\,{T^3}\,\cos ({\it \omega}\,T)  
    \nonumber \\
   & & + \;\;{\it \omega}\,{{\pi }^2}\,{\it \tau_0}\,\cos ({\it \omega}\,T) - 
       {{{\it \omega}}^3}\,{T^2}\,{\it \tau_0}\,\cos ({\it \omega}\,T) - 
       {\it \omega}\,{{\pi }^2}\,T\,
        \cos ({\it \omega}\,\left( T - {\it \tau_0} \right) )  \nonumber \\ 
   & & + \;\; {{{\it \omega}}^3}\,{T^3}\,
        \cos ({\it \omega}\,\left( T - {\it \tau_0} \right) ) + 
       {{\pi }^2}\,\sin ({\it \omega}\,T) - 
       3\,{{{\it \omega}}^2}\,{T^2}\,\sin ({\it \omega}\,T) \nonumber\\
   & & + \;\;  
       {{\pi }^2}\,\sin ({\it \omega}\,
          \left( T - {\it \tau_0} \right) )- 3\,{{{\it \omega}}^2}\,{T^2}\,
        \sin ({\it \omega}\,\left( T - {\it \tau_0} \right) ) - 
       {{\pi }^2}\,\sin ({\it \omega}\,{\it \tau_0}) \nonumber \\
  & & +  \;\;
       3\,{{{\it \omega}}^2}\,{T^2}\,\sin ({\it \omega}\,{\it \tau_0})
        \Bigr)   \;\; .
\eea
(Also in these cases the apparent poles in $b(\tau_0)$ and  $c(\tau_0)$ for
$\omega = \pm \frac{\pi}{T}$ are accompanied by zeros.)
The results of this computations are shown in figure 6 and 7, once more with
no  evidence for bound-state stabilization. A comparison with
the linear switch on and off shows that the ionization probability for
sine-squared  turn-on and offs is lower. The effect is larger for
longer ramps.

\section{Conclusions}

We have investigated the ionization probability for the hydrogen atom
when exposed to ultra-intense shortly pulsed laser radiation of various
types of pulse shapes. In comparison with \cite{FKS}, we  extended our
analysis to the situation which is applicable to any bound-state
$\psi_{nlm}$ and in particular for the $\psi_{100}$-state we carried out
the computation until the end for the stronger upper (\ref{upo})  
and lower (\ref{lowo}) bounds. We overcome the shortcoming of \cite{FKS}
which did not allow definite statements for
pulses of  durations longer than one atomic unit by 
investigating the bounds for higher values of $n$. A direct comparison 
between existing numerical computations for small n, in particular n=1,
and reasonably long pulse durations is at present not feasible. As our
computations show (see also \cite{PontS}) there is of course a quantitative
different behaviour for different values of n. However, qualitatively
we obtain the same behaviour (refer figure 10) and therefore   
we do not think this to be of any physical significance. It would be very 
interesting to carry our analysis further and also investigate the effect 
resulting from varying l and m. In principle our equations already allow
such an analysis, but due to the sum in (A.6) the explicit expressions 
will be rather messy and we will therefore omit them here.

We regard the lack of  support 
for the existence of bound-state stabilization in a realistic three
dimensional atom resulting  from these type
of arguments, even for high values of n, as more convincing than for 
instance the support for stabilization based on  one-dimensional toy models.

For the situation when the total classical momentum transfer 
$b(\tau) $ and the total classical displacement $c(\tau) $ 
are non-vanishing we confirm
once more the results of \cite{FKS} and do not find any
evidence for bound-state stabilization for ultrashort pulses.
This holds for various types of pulses, whether they are switched
on (and off), smoothly or not. We therefore agree with Geltman in the
conclusion that smooth pulses in general will only prolong the onset of
ionization but will not provide a mechanism for stabilization. 

There is however a particular way of switching on and off, such that 
$b(\tau) = 0 $, but $c(\tau) \neq 0 $. These type of pulses are used
for instance in \cite{Grobe,SuEb}. Unfortunately our bounds do not permit
to make any definite statement about this case, since the lower bound is
not applicable (in the sense that then the necessay condition 
$\frac{1}{2}b^2(\tau) > -E$ for th validity of the lower bound is not
fulfilled)
and the upper bound gives for typical values  of the frequency 
and field strength ionization probabilities larger than one. So in
principle for these type of pulses the possibility of bound-state
stabilization remains. It would be very interesting to find alternative
expressions for the upper and lower bound which allow conclusions on 
this case.

For the case $b(\tau) = c(\tau) = 0 $ the upper bound $P_u$ remains an
increasing function of the field strength due to the properties of 
the Hilbert space norm of the difference of the potential in the
Kramers-Henneberger frame and in the laboratory frame applied to 
the state $\psi_{100}$. The weaker upper bound takes on the value 
$P_{uw}(\psi_{n00}) = \frac{4 \tau^2}{n^3}  $, which at first sight
seems counterintuitive, since it implies that the upper bound
decreases with increasing n, i.e. for states close to the ionization 
threshold, and fixed $\tau$. Classically this may, however, be understood
easily. For closed Kepler orbits, i.e. ellipses, with energies sufficiently
close to zero (depending on $\tau$), for any pulse with small 
$b(\tau)$ and $ c(\tau) $, these quantities will be very close to the 
actual changes, caused by the pulse, of the momentum and the 
coordinate, respectively. So in this case ionization, i.e. the transition
to a hyperbolic or parabolic orbit will therefore be very unlikely. This kind
of behaviour was also observed in \cite{PontS} for a Gau\ss ian pulse, for 
which $b(\tau)=0$ and $c(\tau)\neq 0$.
 
\par 
{\bf Acknowledgment:} We would like to thank J.H. Eberly, S. Geltman and  
V. Kostrykin for very useful discussions and correspondences. One of the
authors (CFMF) is  supported by the DAAD.

\appendix
\section*{Appendix}
\renewcommand{\theequation}{\mbox{{A}.\arabic{equation}}}
\setcounter{equation}{0} In this appendix we will provide the explicit
calculation of the term 
\beq\label{A1} 
N^2(\vec{y},\psi)\;= \;\langle
\psi,V(\vec{x})^2 \psi \rangle + \langle \psi,V(\vec{x} -\vec{y})^2
\psi\rangle-2 \langle\psi, V(\vec{x} -\vec{y}) V(\vec{x})\psi \rangle
\eeq 
For $ \psi = \psi_{nlm}$ the first term is well known to equal 
$\frac{1}{n^3(l + 1/2)}$
\cite{BSLL}.  We did not find a computation for the matrix element 
involving the Coulomb potential in the Kramers-Henneberger frame 
in the literature and will therefore present it
here. Starting with the familiar  expansion of  the shifted Coulomb
potential in terms of spherical harmonics 
\beq
\label{entwick}
{1\over |\vec{x}-\vec{y}|} =  \sum^\infty_{l =0} 
\left( \frac{ r_<^l}{  r_>^{l+1} } \right)\sqrt{ \frac{ 4 \pi}{2l +1} }
Y_{l \; 0} (\vartheta,\phi)
\eeq
where $r_< = Min ( |\vec{x}|, |\vec{y}| )$ and  
$r_> = Max ( |\vec{x}|, |\vec{y}| )$, 
we obtain
\bea
\langle \Psi_{nlm} |\; |\vec{x} - \vec{y}|^{-1} \; |\vec{x}|^{-1} |\Psi_{n l m} \rangle & = & \sum_{l' =0}^{\infty} \int d\Omega
Y_{l \; m}^* Y_{l ' \; 0} Y_{l \; m} 
\sqrt{ \frac{4 \pi}{2 l' +1} } \nonumber \\& &\left( \int\limits_0^{ |\vec{y}|} dr \left( \frac{r}{ |\vec{y}|}\right)^{l' +1} R_{nl}^2 +    \int\limits_{ |\vec{y}|}^{\infty} dr \left( \frac{ |\vec{y}|}{r} \right)^{l'} R_{nl}^2 \right) 
\nonumber\eea
which by the well known formula from angular momentum theory 
\beq \int d\Omega Y_{l \; m}^* Y_{l_1 \; m_1} Y_{l_2 \; m_2} \;=\;\sqrt{ \frac{(2 l_1 +1)(2 l_2 +1) }{ 4 \pi (2 l +1)}  }
\langle l_1 l_2 ; 00 | l 0 \rangle
\langle l_1 l_2 ; m_1 m_2 | l m \rangle \label{add}
\eeq
leads to
\beq \label{eq: A4}
\sum_{ l' =0}^{\infty} \langle l l' ; 00 | l 0 \rangle
\langle l l' ; m0 | l m \rangle \;\;
\left( \int\limits_0^{ |\vec{y}|} dr \left( \frac{r}{ |\vec{y}|}\right)^{l' +1} R_{nl}^2 +    \int\limits_{ |\vec{y}|}^{\infty} dr \left( \frac{ |\vec{y}|}{r} \right)^{l'} R_{nl}^2 \right)\; .
\eeq
Here $\langle l_1 l_2; m_1 m_2  | l m \rangle$ denote the Wigner or 
Clebsch-Gordan coefficients in the usual conventions 
(see e.g. \cite{Inui}). 

We shall now consider the term
\beq
\langle\Psi_{nlm}|\;|\vec{x}-\vec{y}|^{-2}\; | \Psi_{n l m} \rangle \;\;
\label{quad}
\eeq
Employing (\ref{entwick}) and the formula 
$$
Y_{l_1 m_1 } Y_{l_2 m_2 }= \sqrt{\frac{(2l_1+1)(2l_2+1)}{4\pi}}
\sum_{ l' m'} \frac{1}{(2l'+1)} Y_{l' m' }
\langle l_1 l_2; m_1 m_2  | l' m' \rangle
\langle l_1 l_2; 00  | l' 0 \rangle
$$
yields
\beq
\frac{1}{|\vec{x} - \vec{y}|^2 } = \sum_{k,l'} 
\sum_{\tilde{l}= |k - l'|}^{ |k + l'|}
\frac{r_<^{k + l'}}{r_>^{k + l'+2 }} \;
\sqrt{\frac{4 \pi }{2 \tilde{l } +1  } } \;
\langle k l' ; 00  | \tilde{l} 0 \rangle^2 \;
Y_{\tilde{l} 0 } \;\; .
\eeq
Once again applying (\ref{add}) shows that (\ref{quad}) equals
\beq  \label{A8}
\sum_{\tilde{l},\bar{l}, l'}
\langle \tilde{l} l' ; 00 | \tilde{l} 0 \rangle^2
\langle \bar{l} l ; 0m | l m \rangle
\langle \bar{l} l ; 00 | l 0 \rangle    \;\;
\left( \int\limits_0^{ |\vec{y}|} dr \left( \frac{ r }
{ |\vec{y}|} \right)^{l' + \tilde{l} +2} R_{nl}^2 +    
\int\limits_{ |\vec{y}|}^{\infty} dr \left( \frac{ |\vec{y}|}{r} 
\right)^{l' + \tilde{l} } R_{nl}^2 \right)\; .
\eeq
For s-states, i.e. ($l=0$), we may carry out the sums over 
the Clebsch-Gordan coefficients easily. In (\ref{eq: A4}) the only 
contribution comes from $ l' =0$ and we trivially obtain
\beq 
\langle \Psi_{n00} |\; |\vec{x} - \vec{y}|^{-1} \; |\vec{x}|^{-1} 
|\Psi_{n00} \rangle \;\; = \;\;
\int\limits_0^{ |\vec{y}|} dr  \frac{r}{ 
|\vec{y}|} R_{n0}^2 + \int\limits_{ |\vec{y}|}^{\infty} 
dr R_{n0}^2 \; .
\eeq
In  (\ref{A8}) the sum over $\bar{l}$  contributes only for
$\bar{l} = 0$ and together with $ \langle \tilde{l} l' ; 00 | 00 \rangle^2 $
$=  \frac{ \delta_{\tilde{l}l'}}{2 \tilde{l} + 1}$ it leads to 
\beq \label{eq: squ} 
\langle\Psi_{n00}|\;|\vec{x}-\vec{y}|^{-2}\; | \Psi_{n00} \rangle \;\;
= \;\;
\sum_{l=0 }^{\infty} \frac{1}{2 l + 1}
\left( \int\limits_0^{ |\vec{y}|} dr \left( \frac{ r }
{ |\vec{y}|} \right)^{2 l +2} R_{nl}^2 +    
\int\limits_{ |\vec{y}|}^{\infty} dr \left( \frac{ |\vec{y}|}{r} 
\right)^{ 2 l } R_{nl}^2 \right)\; .
\eeq
We turn to the case $n=1$  (with $ \Psi_{100} = 
\frac{2}{\sqrt{4 \pi}}e^{- |\vec{x}|}$)   
for which  (\ref{eq: A4}) becomes
\beq \label{A5}
\langle \Psi_{100} |\; |\vec{x} - \vec{y}|^{-1} \; |\vec{x}|^{-1} 
|\Psi_{100} \rangle \; = \; \frac{ 1 - e^{-2 |\vec{y}|}}{|\vec{y}| } 
 \;\; .
\eeq
As consistency check one may consider the asymptotic behaviours 
$|\vec{y}| \rightarrow \infty$ and $|\vec{y}| \rightarrow 0$, which
give, as expected, 0 and 2 respectively. Using the series expansion for 
the logarithm,  (\ref{eq: squ}) for $n=1$ becomes 
\beq
\langle \Psi_{100} |\; |\vec{x} - \vec{y}|^{-2} \; |\Psi_{100} \rangle \; = \; \frac{2 }{| \vec{y}|} \left(
 \int\limits^{ |\vec{y}|}_0 dr \ln 
\left( \frac{| \vec{y}| +r  }{| \vec{y}| -r  } \right)
r e^{- 2 r  } +
 \int\limits_{ |\vec{y}|}^{\infty } dr \ln 
\left(  \frac{r + | \vec{y}|  }{r - | \vec{y}|  } \right)
r e^{- 2 r  } \right)\; .
\eeq
Using then the integrals
\bea
\int dr \ln(1 \pm r) r e^{-2 c r }  &=& \frac{1}{4 c^2 } \Biggl(
(1 \mp 2c) e^{ \pm 2 c r } Ei\left( \mp 2 c (1\pm r) \right) \nonumber  \\
 & & - e^{- 2 c r } \left(1 + (1 +2cr) \ln(1\pm r) \right)\Biggr)\\ 
\int dr \ln(1 \pm r^{-1}) r e^{-2 c r }  &=& \frac{1}{4 c^2 } \Biggl(
(1 \mp 2c) e^{ \pm 2 c r } Ei\left( 2 c ( \mp 1 - r) \right) \nonumber  \\
 & &  - Ei\left(- 2 c  r \right) - (1 + 2cr) e^{ - 2 c r }
\ln(1\pm r^{-1}) \Biggr) \;\;\;\;\;
\eea
we obtain
\beq \label{A12}
\langle \Psi_{100} |\; |\vec{x} - \vec{y}/2|^{-2} \; |\Psi_{100} \rangle \; = ( 1 - |\vec{y}|^{-1}) e^{-|\vec{y}|)}
Ei(|\vec{y}|) +  ( 1 - |\vec{y}|^{-1}) e^{|\vec{y}|)}
Ei(-|\vec{y}|)
\eeq
As a consistency check we may again consider the asymptotic behaviour, that
is $|\vec{y}| \rightarrow 0$ and  $|\vec{y}| \rightarrow \infty$, which 
gives correctly 2 and 0, respectively. Assembling now (\ref{A1}), (\ref{A5})
and  (\ref{A12}) gives as claimed (\ref{HNorm}). In the same fashion one
may also compute $N(\vec{y},\psi_{nlm})$ for arbitrary $n,l$ and $m$.


\begin{figure}[t]
\begin{caption}
{The Hilbert space 
norm of the difference of the potential in the Kramers-Henneberger frame
and in the laboratory frame applied to the state $\psi_{100}$  versus
the classical displacement c.}
\end{caption}
\label{fig1}

\begin{caption}
{Upper (three curves on the left) and lower bound 
($P_l$ and $P_u$) for the ionization 
probability of the $\psi_{100}$-state through a static laser field 
$E_0$. The dotted line 
corresponds to $E_0=5$ a.u., the dashed line to $E_0=10$ a.u. and the solid
line to $E_0=20$ a.u. The time is in a.u.}
\end{caption}
\label{fig2}

\begin{caption}
{Upper (three curves on the left) and lower bound  ($P_l$ and $P_u$)
for the ionization 
probability of the $\psi_{100}$-state through a linearly polarized 
monochromatic laser field 
$E(t) = E_0 \sin(\omega t) $; $\omega= 1.5 $ a.u. The dotted line 
corresponds to $E_0=5$ a.u., the dashed line to $E_0=10$ a.u. and the solid
line to $E_0=20$ a.u. The time is in a.u.}
\end{caption}
\label{fig3}

\begin{caption}
{Lower bound  ($P_{lw}$) for the ionization 
probability of the $\psi_{10 \; 00}$-state through a linearly polarized 
monochromatic laser field 
$E(t) = E_0 \sin(\omega t) $, $E_0= 2$ a.u. The dotted line 
corresponds to $\omega = 0.4$ a.u. and the solid
line to $ \omega = 4$ a.u. The time is in a.u.}
\end{caption}
\label{fig4}

\begin{caption}
{Lower bound for the ionization ($P_{lw}$)
probability of the $\psi_{20 \;00}$-state through a linearly polarized 
monochromatic laser field 
$E(t) = E_0 \sin(\omega t) $, $\omega= 1.5$ a.u.,  $E_0= 20$ a.u. 
The time is in a.u.}
\end{caption}
\label{fig5}

\begin{caption}
{Lower bound ($P_{lw}$)  for the ionization 
probability of the $\psi_{34 \;00}$-state through a linearly polarized 
monochromatic laser field with a trapezoidal and a sine-squared turn-on
and turn-off enveloping function, upper and lower curve of the same line
type, respectively. 
(solid line: $\frac{5}{4}-12-\frac{5}{4}$ pulse,
dashed line: $\frac{9}{4}-10-\frac{9}{4}$ pulse and
dotted line: $\frac{17}{4}-6-\frac{17}{4}$ pulse), 
$\omega= 1.5$ a.u. }
\end{caption}
\label{fig6}
\end{figure}
\begin{figure}[t]
\begin{caption}
{Upper bound ($P_{lw}$) for the ionization 
probability of the $\psi_{34 \;00}$-state through a linearly polarized 
monochromatic laser field with a trapezoidal and a sine-squared turn-on
and turn-off enveloping function, upper and lower curve of the same line
type, respectively. 
(solid line: $\frac{1}{2}-6-\frac{1}{2}$ pulse,
dashed line: $\frac{3}{2}-4-\frac{3}{2}$ pulse and
dotted line: $\frac{5}{2}-2-\frac{5}{2}$ pulse), 
$\omega= 1.5$ a.u. }
\end{caption}
\label{fig7}

\begin{caption}
{Lower bound ($P_{lw}$) for the ionization 
probability of the $\psi_{30 \;00}$-state through a linearly polarized 
monochromatic laser field with a sine-squared enveloping function
$E(t) = E_0 \sin(\omega t) \sin(\Omega t)^2    $, $\omega= 0.2$ a.u.,  
$\Omega= 0.01$ a.u., $E_0= 20$ a.u. The time is in a.u. }
\end{caption}
\label{fig8}

\begin{caption}
{Lower bound ($P_{lw}$) for the ionization 
probability of the $\psi_{30 \; 00}$-state through a linearly polarized 
monochromatic laser field with a sine-squared enveloping function
$E(t) = E_0 \sin(\omega t) \sin(\Omega t)^2    $, $\omega= 0.2$ a.u.,  
$\Omega= 0.01$ a.u. The dotted line 
corresponds to $E_0=5$ a.u., the dashed line to $E_0=10$ a.u. 
and the solid
line to $E_0=20$ a.u. The time is in a.u.}
\end{caption}
\label{fig9}

\begin{caption}
{Lower bound ($P_{lw}$) for the ionization 
probability of the $\psi_{n00}$-state through a linearly polarized 
monochromatic laser field with a sine-squared enveloping function
$E(t) = E_0 \sin(\omega t) \sin(\Omega t)^2    $, $\omega= 0.8$ a.u.,  
$\Omega= \omega/13.5$ a.u. The dotted line 
corresponds to $n=40$, the dashed line to $n=35$  
and the solid
line to $n=30$.}
\end{caption}
\label{fig10}

\end{figure}


\begin{thebibliography}{99}

\bibitem{Lam} P. Lambropoulos, \PRL {\bf 29} (1972) 453.
\bibitem{Wil} W. Rudolph and B. Wilhelmi, {\em Light Pulse Compression}
(Harwood Acad. Publ., London, 1989) 
\bibitem{GordonVolk} W. Gordon {\em Zeit. f\"ur Physik} {\bf 40} (1926)
117; D.M. Volkov, {\em Zeit. f\"ur Physik} {\bf 94} (1935) 250. 
\bibitem{Keld} L.V. Keldysh, \SOV {\bf 20} (1965) 1307;
 A.M. Perelomov, V.S. Popov and M.V. Terentev; 
\SOV {\bf 23} (1966) 924; \SOV {\bf 24} (1967) 207;
F.H.M. Faisal, {\em Jour. of Phys.} {\bf B6} (1973) L89. 
\bibitem{GeltT} S. Geltman and M.R. Teague, {\em Jour. of Phys.} {\bf B7}
(1974) L22.
\bibitem{Reiss}  H.R. Reiss, \PRA {\bf A22} (1980) 1786.
\bibitem{Gelt1} S. Geltman, \PRA {\bf A45} (1992) 5293.
\bibitem{Grobe}R. Grobe and M.V. Fedorov, \PRL {\bf 68} 
(1992) 2592; R. Grobe and M.V. Fedorov, \JPB {\bf B26} (1993) 1181.
\bibitem{Gelt2} S. Geltman, \JPB {\bf B27} (1994) 257.
\bibitem{Gelt3} S. Geltman, \JPB {\bf B27} (1994) 1497. 
\bibitem{Col} L.A. Collins and A.L. Merts, \PRA {\bf A37} (1988) 2415.
\bibitem{Bard} J.N. Bardsley, A. Sz\"oke and M.J. Comella, \JPB {\bf B21}
(1988) 3899.
\bibitem{LaG} K.J. LaGattuta, \PRA {\bf A40} (1989) 683.
\bibitem{SuEb}  J. Javanainen, J.H. Eberly and
Q. Su, \PRA {\bf A38} (1988) 3430;\PRL {\bf 64} (1990) 862; 
Q. Su and J.H. Eberly,
{\em Jour. of Opt. Soc. Am.} {\bf 7} (1990) 564
Q. Su and J.H. Eberly   {\em J. Opt. Soc. Am.} {\bf B7} (1990) 564;
C.K. Law, Q. Su and J.H. Eberly, \PRA {\bf 44} (1991) 7844;
Q. Su and J.H. Eberly,
\PRA {\bf A43} (1991) 2474;
Q. Su   {\em Laser Phys.} {\bf 2} (1993) 241;
Q. Su, B.P. Irving, C.W. Johnson and J.H. Eberly, 
\JPB {\bf B29} (1996) 5755.
\bibitem{Burnett} 
V.C. Reed and K. Burnett, \PRA {\bf A42} (1990) 3152; 
\PRA {\bf A43} (1991) 6217
\bibitem{Dorr}
O. Latinne, C.J. Joachain and M. D\"orr,
{\em Europhys. Lett.} {\bf 26} (1994) 333; M. D\"orr, O. Latinne 
and C.J. Joachain, \PRA {\bf 52} (1995) 4289
\bibitem{Floquet}  J.H. Shirley, \PRA {\bf B138} (1965) 979.
\bibitem{Pot} R.M. Potvliege and R. Shakeshaft, \PRA {\bf A38} (1988) 1098;   
\PRA {\bf A38} (1988) 4597; \PRA {\bf A38} (1988) 6190;  
\PRA {\bf A40} (1990) 4061; 
M. D\"orr, R.M. Potvliege, D. Proulx and R. Shakeshaft,
\PRA {\bf A43} (1991) 3729; M. D\"orr, P.G. Burke, C.J. Joachain, C.J. Noble,
J. Purvis and M. Terao-Dunseath, \JPB {\bf B25} (1993) L275.
\bibitem{Faisal}
L. Dimou and F.H.M. Faisal, {\em Acta Physica Polonica}
{\bf A86} (1993) 201; L. Dimou and F.H.M. Faisal, \PRA {\bf A46} (1994) 4564;
F.H.M. Faisal, L. Dimou, H.-J. Stiemke and M. Nurhuda, {\em Journ. of
Nonlinear Optical Physics and Materials} {\bf 4} (1995) 701.
\bibitem{HFA} M. Gavrila  and J.Z. Kaminski \PRL {\bf 52} (1984) 613;
 M.J. Offerhaus, J.Z. Kaminski and M. Gavrila, \PL 
{\bf 112A} (1985) 151;  M. Gavrila,  M.J. Offerhaus and J.Z. Kaminski, 
\PL {\bf 118A} (1986) 331; M. Pont, M.J. Offerhaus  and M. Gavrila, 
{\em Z. Phys.} {\bf D9} (1988) 297; J. van de Ree, J.Z. Kaminski and 
M. Gavrila, \PRA  {\bf A37} (1988) 4536; M. Pont, N.R. Walet, M. Gavrila
and C.W. McCurdy; \PRL {\bf 61} (1988) 939;
M. Pont and M. Gavrila, \PRL {\bf 65} (1990) 2362.
\bibitem{classical} J.G. Leopold and I.C. Parcival, \PRL {\bf 41} (1978)
944; J. Grochmalicki, M. Lewenstein and K. Rza\.zewski, \PRL {\bf 66}
(1991) 1038; R. Grobe and C.K. Law, \PRA {\bf A44} (1991) 4114;
F. Benvenuto, G. Casati and D.L. Stepelyanski, \PRA {\bf A45} (1992) R7670; 
S. Menis, R. Taieb, V. Veniard and A. Maquet,
\JPB {\bf B25} (1992) L263. 
\bibitem{BRK} K. Burnett, V.C. Reed and P.L. Knight, \JPB {\bf 26}
(1993) 561.
\bibitem{KulS} J.H. Eberly and K.C. Kulander, {\em Science} {\bf 262} 
(1993) 1229.
\bibitem{KrainDel}
N.B. Delone and V.P. Krainov, {\em Multiphoton Processes in Atoms}
(Springer Verlag, Berlin, 1994) Chapter 10.
\bibitem{Gelt4} S. Geltman, {\em Chem. Phys. Lett.} {\bf 237} (1995)
286.
\bibitem{Chen} Q. Chen and I.B. Bernstein, \PRA {\bf A47} (1993) 4099. 
\bibitem{Krain} V.P. Krainov and M.A. Preobrazenski, 
\SOV {\bf 76} (1993) 559.
\bibitem{expe} M.P. de Boer, J.H. Hoogenraad, R.B. Vrijen, 
L.D. Noordam and H.G. Muller, \PRL (1993) {\bf 71} 3263; 
M.P. de Boer, J.H. Hoogenraad, R.B. Vrijen and  L.D. Noordam, \PRA
{\bf A50} (1994) 4133. 
\bibitem{EKS} V. Enss, V. Kostrykin and R. Schrader, 
\PRA {\bf A50} (1995) 1578.
\bibitem{KS1} V. Kostrykin and R. Schrader, \JPB {\bf B28} (1995) L87.
\bibitem{KS2}  V. Kostrykin and R. Schrader, {\em Ionization of Atoms 
and Molecules by
short strong laser pulses}, Sfb 288, preprint 185, Berlin (1995),
submitted to \JPB {\bf A}.
\bibitem{FKS}  A. Fring, V. Kostrykin and R. Schrader, \JPB 
{\bf B29} (1996) 5651.
\bibitem{BSLL} H.A. Bethe and E.E. Salpeter, {\em Quantum Mechanics
of One and Two-Electron Atoms} (Springer, Berlin, 1957),
L.D. Landau and E.M. Lifschitz, {\em Quantum Mechanics}
(Pergamon Press, New York, 1977).
\bibitem{CFKS} H.L  Cycon, R.G. Froese, W. Kirsch, W. and B. Simon,  
{\em Schr\"odinger Operators} (Springer, Berlin, 1987).
\bibitem{RS} M. Reed and B. Simon, {\em  Methods of Modern Mathematical 
Physics} (Academic Press, New York, Vol. 2, 1972).
\bibitem{K} H.A. Kramers, {\em  Collected Scientific Papers}, 
(North-Holland, Amsterdam, 1956).
\bibitem{H} W.C. Henneberger, \PRL {\bf 21} (1968) 838.
\bibitem{PontS} M. Pont and R. Shakeshaft, \PRA {\bf A44} (1991) R4110. 
\bibitem{Inui} Y. Inui, Y. Tanabe and Y. Onodera, {\em Group Theory
and its Application in Physics} (Springer, Berlin, 1996).

\end{thebibliography}
\end{document}